\documentstyle[aps,prl,multicol]{revtex}

\newcommand{\beq}{\begin{eqnarray}}
\newcommand{\eeq}{\end{eqnarray}}
\newcommand{\n}{\nonumber}

\newcommand{\ep}{\epsilon_{k,\omega}}

\newcommand{\bleq}{\ifpreprintsty
                   \else
                   \end{multicols}\vspace*{-3.5ex}{\tiny 
                   \noindent\begin{tabular}[t]{c|}
                   \parbox{0.493\hsize}{~} \\ \hline \end{tabular}}
                   \fi}
\newcommand{\eleq}{\ifpreprintsty
                   \else
                   {\tiny\hspace*{\fill}\begin{tabular}[t]{|c}\hline
                    \parbox{0.49\hsize}{~} \\ 
                    \end{tabular}}\vspace*{-2.5ex}\begin{multicols}{2}
                    \fi}

\begin{document}

\title
{  
Interlayer c-axis transport in the normal state of cuprates}

\author{ Misha Turlakov and  Anthony J. Leggett}
\address{ Department of Physics, University of Illinois at Urbana-Champaign}
\date{\today}
\maketitle

\begin{abstract}
A theoretical model of c-axis transport properties in cuprates is proposed. 
Inter-plane and in-plane charge fluctuations make hopping
between planes incoherent and diffusive (the in-plane momentum is not
conserved after  tunneling). 
The non-Drude optical conductivity $\sigma_c(\omega)$
and the power-law temperature dependence of the {\it dc} conductivity
are generically explained by the strong fluctuations excited in the process
of tunneling. Several microscopic models of the charge fluctuation spectrum
are considered.

% Microscopic charge fluctuations are estimated for several models.
% Other experimental observations are discussed in the framework of this model.

\end{abstract}
\vskip2pc

\tighten
% \narrowtext

\begin{multicols}{2}

% Introduction
% Layered and strongly correlated nature of cuprates is an essential part of
% high-temperature superconductivity puzzle.

Despite the strongly two-dimensional layered structure of the high-temperature 
cuprate superconductors, features associated with the third dimension, perpendicular
to the $CuO_2$ planes, may be an important ingredient in their superconductivity.
In fact, it is well accepted that a certain degree of Josephson-type coupling between 
different planes is necessary to suppress the two-dimensional fluctuations, which will
otherwise destroy the superconducting long-range order. However, the systematic dependence
of the critical temperature $T_c$ on the number of layers in the unit cell 
(together with the absence of evidence for
strong fluctuations effects above $T_c$ at optimal doping, which suggests that these 
 fluctuations are not the major reason for this systematic dependence) points
almost unambiguously to the conclusion that theories formulated for a single plane cannot
be the whole story. Either hopping between planes\cite{Anderson}, or Coulomb interaction
between them\cite{Leggett}, or both, is an important factor in raising the critical
temperature (and, perhaps, in some cases also for lowering it, see Ref. \cite{Leggett}).
In the light of this, the study of the c-axis optical and transport properties is more than
just a minor diversion from the main issue. 

% In order to answer this question
% we need to analyze  c-axis optical and transport properties.

These c-axis optical and transport properties are very puzzling and anomalous\cite{cooper}.
Most remarkable is the fact that the temperature dependence of the {\it dc} c-axis resistivity
$\rho_c (T)$, in  sharp contrast to the well-known linear $T$-dependence of 
the in-plane resistivity
$\rho_{ab}(T)$, is non-universal, being described in most cases by
 a power law $\rho_c(T) \sim T^\gamma$,
 where however the exponent $\gamma$ can be anything in between approximately $+1$ and $-1$. 
%In the literature this behavior is sometimes characterized as ``metallic'' or ``semiconducting''
%corresponding to $\gamma > 0$ or $\gamma < 0$, but we prefer to avoid this somewhat tendentious language.
The optical conductivity $\sigma_c (\omega)$ is roughly frequency 
independent from low frequencies
up to mid-infrared frequencies (except in the case of some overdoped cuprates 
($YBa_2Cu_3O_7$ and $La_{2-x}Sr_xCu O_4$));
 the numerical value is below the Mott-Ioffe-Regel minimum metallic conductivity.
This behavior, which is dramatically different from the behavior of the in-plane resistivity,
has been christened ``confinement''\cite{Anderson}. Thus, despite the dramatic differences
in the raw data between the different cuprate families, one can isolate at least two elements
which  may be legitimately called ``universal'': a non-Drude optical conductivity of a magnitude
below the Mott limit, and a power-law temperature dependence of the {\it dc} 
resistivity $\rho_c(T)$ (albeit with
a material-specific exponent).
In this paper we develop a framework for the explanation of these universalities, which will
hopefully shed light on the role of the inter-plane and in-plane Coulomb interaction as well
as on the more obvious one of the inter-layer hopping.

% ?? summary of results ? not necessary- too much repetition
% brief review of experiments

%A natural starting point for the study of the c-axis transport properties is the assumption that the process
%of transport of an electron, certainly between different unit cells and arguably even between different 
%planes within a unit cell, is of diffusive nature on the scale of a lattice constant (*). The strongest 
%qualitative argument is the form and the value of the non-Drude optical conductivity:
%(*) to fit the observed conductivity to a conventional band picture would require either a 
%mean free path shorter than the c-axis lattice spacing or a Fermi surface with a very tiny dispersion
%along the c-axis direction. However, the latter mechanism cannot easily explain the non-Drude
%frequency dependence.

On the basis of the above experimental observations, we can make the assumption 
of incoherent transport
(diffusive tunneling) in the c-direction. The inter-plane (or rather inter-unit cell) hopping
time $\tau_{hop}$ can be estimated from the {\it dc} c-axis resistivity. Relating the diffusion
constant $D$ to the hopping time $\tau_{hop}$ by $D=d^2/\tau_{hop}$, 
% (assuming the out-of-plane mean free path
% equal to the inter-unit cell distance $d$), 
we can derive a model-independent relation
between the conductivity $\sigma_c$ and the hopping time: 
$ \sigma_c=e^2 \nu_{2d} \frac{d}{\tau_{hop}} $, where $\nu_{2d}$ is a two-dimensional 
density of states.  
Using this formula  and the experimental values for the c-axis conductivity we can estimate
the c-axis hopping time $\tau_{hop}$. The strong two-dimensionality of 
the electron motion becomes obvious
if we compare the hopping time $\tau_{hop}$ with the in-plane scattering time $\tau_{ab}$. 
As is well known, the in-plane scattering time $\tau_{ab}$ is of order of $\hbar/kT$.
Direct comparison \cite{hop} shows that  for most materials the c-axis hopping time is 
{\it much longer} than  the scattering time in the plane. These two times are comparable
only for overdoped $YBa_2Cu_3O_7$ and $La_{2-x}Sr_xCu O_4$ suggesting a crossover to a different
regime of c-axis transport; this is confirmed by the experimental observation 
of the Drude-like frequency dependence
of the conductivity $\sigma_c (\omega)$ for these compounds. 

%This fact allows us to treat the motion and scattering in the planes as {\it fast}
%relative to the tunneling between planes (*). 

Many approaches have been suggested\cite{Anderson,cooper} to describe the c-axis
transport properties. Most of them stem from phenomenologically 
assumed  in-plane Green functions. One remarkable example is 
a non-Fermi ``Luttinger'' liquid  theory which
explains naturally the ``confinement'' for c-axis motion in the normal state
\cite{Anderson}. Many other theories are essentially based on the Fermi liquid theory
modified by strong correlations\cite{cooper,Millis}.  

We take a quite different approach to the problem.
Although in our approach the in-plane motion (expressed by in-plane Green functions) 
is undeniably 
important, we show
that most of the c-axis properties can be qualitatively understood 
 on the basis of knowledge of the spectrum
of in-plane and inter-plane {\it charge density fluctuations  excited during the process of the
inter-plane tunneling}. 
The spectrum of charge fluctuations can be directly
measured experimentally  by  optical reflectivity and electron-energy-loss spectroscopy (EELS).
%This remark is especially important in view of the current lack of 
%an agreed understanding of the in-plane
%normal state properties.

The essential physical picture of our approach is that the c-axis tunneling
 is strongly suppressed
by charge fluctuations excited in the process of tunneling\cite{Leggett-bras}.
 This anomaly (the so called
Coulomb blockade) is widely observed in many strongly correlated
and mesoscopic systems. 
Other examples of this class of phenomena are orthogonality catastrophes and
a zero-bias anomaly in diffusive systems.
The ubiquity of the Coulomb blockade phenomena (static or dynamic) in correlated systems
 indicates 
that the anomalous c-axis transport properties may be merely  
a consequence of strong correlations in the cuprates.
The necessary condition for the appearance of the Coulomb blockade phenomenon is
the strong effective coupling of the tunneling electron to the collective
excitations of the liquid.
In fact, we can think about the Coulomb blockade as
a ``high-energy'' phenomenon of order of Coulomb energy per electron and  independent 
of the low-energy quasiparticle spectral properties, be they fermi- or non-fermi liquid.
In other words, the tunneling electron couples to the excitations in the broad
range of frequencies from low up to high frequencies.
The non-fermi liquid property (property not present for a three-dimensional 
Fermi liquid
or for a electron gas in the RPA approximation) which is responsible for the anomalous
c-axis properties is simply the large density of ``detuning'' 
charge fluctuations (as observed
by optical and Raman spectroscopy) over a broad frequency range up to 
mid-infrared frequencies.

Empirically, an  overview of the in-plane conductivity and the c-axis conductivity
 in various families
of cuprates does not reveal any obvious correlation between their temperature dependencies. 
On the other hand, the perturbative diagrammatic expression 
(assuming the equivalent Green's functions in each plane and uncorrelated impurities)
for the in-plane conductivity and out-of-plane conductivity\cite{Mahan}
are equivalent up to the vertex functions. 
%of the latter contain 
%the tunneling matrix element $t_\perp$ instead of the fermi momentum for the former.
Therefore the difference
in the temperature and frequency  dependence of the in- and out- plane conductivities 
must come {\it exclusively} from the  interplane tunneling probability. For instance,
the notion of the ``two-dimensional Luttinger'' liquid is not sufficient by itself
to explain {\it the difference} between the in-plane and out-of-plane resistivities
\cite{Georges}.

One particular approach to explaining the difference between the resistivities 
along the different directions is based on the highly anisotropic form of the
tunneling matrix element $t_\perp (k_x,k_y)$ as a function of the in-plane momentum.
Several authors\cite{Millis} developed a phenomenological approach
(assuming as well a strong anisotropy of quasiparticle lifetimes and the density of states
around the fermi surface) which seem to fit successfully the experimental data.
Two remarks are in order.
First, this approach assumes that the tunneling conserves the in-plane momentum.
This may be the case
in certain situations (possibly, in the superconducting state and in the overdoped regime),
but in general this assumption deserves a close scrutiny by experiment and theory.
In fact, we will argue that the in-plane momentum is not conserved when 
$\tau_{hop}\gg \tau_{ab}$.
% (some other requirements?). 
%In this case, the anisotropy
%of the matrix element $t_\perp (k_x,k_y)$ cannot provide an explanation
%for different in-plane and out-of-plane resistivities. 
Second, the anisotropy
of the tunneling matrix element $t_\perp (k_x,k_y)$ is different for some variations
of cuprates\cite{Xiang}, and , in general, it can be doping dependent.
Thus in some cases, $t_\perp (k_x,k_y)$ may not vanish along the diagonals of the Brillouin
zone making the contribution of the diagonal quasiparticles ($k_x=\pm k_y$) to the
c-axis conductivity non-vanishing contrary to the assumptions of references\cite{Millis}.

In the rest of the paper, we begin by generalizing the standard tunneling formalism,
introducing a non-trivial tunneling probability which accounts for the inelastic
and elastic (momentum scattering) processes. Then, we calculate this tunneling probability
from assumed spectra of ``the detuning fluctuations''. After that, we calculate
the experimentally measured optical conductivity $\sigma_c (\omega, T)$ and the
tunneling conductance $\sigma_c (V)$. Finally, we discuss the complex experimental
situation and possible extensions of the proposed theory. 
In the appendix we discuss several
microscopic models giving the spectrum of the charge fluctuations.

% Objectives:
% Without going into discussion of the nature of normal state (Fermi or Luttinger liquid)
% the goal is to understand c-axis transport properties.

% Theory

{\it Theory. The tunneling formalism.} The tunneling of an electron from one plane 
to another plane can be considered by
using the time-dependent tunneling hamiltonian formalism. The ``blockade effect'' 
due to the excitation
of the electromagnetic modes is accounted by the modulation of the tunneling matrix element by
the Coulomb interaction.
Thus the part of the hamiltonian responsible for the transfer of electrons between planes is:
\beq
&&H_c= \sum_{r_1,r_2} t_{\perp}(r_1,r_2;t) ( a^+_1 (r_1) a_2 (r_2) +  a^+_2 (r_2) a_1 (r_1) ),
\eeq
where the quasi-classical tunneling matrix element $t_{\perp}(r_1,r_2;t)$ is equal to
$t_{\perp} (r_1,r_2) exp \left(-\frac{ie}{c} \int_{r_1}^{r_2} A(z,t) dz \right).$
Gauge invariance dictates the presence of the phase factor 
$\varphi (r_1,t)=(e/c)\int_{r_1}^{r_2}A(z,t) dz$, where the integral is taken 
over a path connecting
two points $r_1$ and $r_2$ on the different planes. 
Since the optimal tunneling trajectory is perpendicular to the planes (along the $z$ axis),
the tunneling matrix element can be written as
$t_{\perp} (r_1,r_2)=t_{\perp} \delta (r_1 -r_2) 
exp \left(-\frac{ie}{c} \int_{z_1}^{z_2} A(z,t) dz \right)$. 
It conserves the in-plane momentum. 
Using the tunneling hamiltonian formalism \cite{Mahan,Barone}, we get the following expression
for the tunneling current $I(t)$ between two planes:

\bleq
\begin{eqnarray}
&&I(t)=-\frac{2e}{\hbar^2} Re \int\int drdr' \int_{-\infty}^{+\infty} 
dt' e^{-i\frac{eVt}{\hbar}}|t_{\perp}|^2
 P(r-r',t-t') S(r-r',t-t'),~~~ 
P(r-r',t-t') \equiv \left< e^{i\varphi (r,t)} e^{-i\varphi (r',t')} \right>
 \label{eq:current}, \\ 
&&S(r-r',t-t' ) \equiv \Theta (t-t' ) 
< G^<_R (r-r',t - t' )G^>_L (r'-r,t'-t) - G^<_L (r-r', t-t') G^>_R (r'-r,t' -t) >  \n 
% &&P_{r-r'}(t-t') \equiv T_r T_{r'}^* \left< e^{i\varphi (r,t)} e^{-i\varphi (r',t')} \right>,
\end{eqnarray}
\eleq\noindent
where $P(r-r',t-t')$ is a phase-phase correlation function
 between two planes  averaged  over the equilibrium fluctuations, and
$V$ is an applied voltage.
The definitions of Green's functions and essential details of the derivation
can be found in the Ref.\cite{Barone}. 
The fact that the hopping time $\tau_{hop}$ is much longer than 
the in-plane scattering time $\tau_{ab}$
allows us to separate the in-plane propagation $S(r-r',t-t')$ and 
the tunneling probability $P(r-r',t-t')$.
It is important to remark at this stage that the long-wavelength fluctuation modes 
(with the wavelength much longer than the in-plane mean free path) can suppress 
the tunneling probability
without effecting the in-plane motion. The scattering by the short-wavelength
fluctuations is accounted by the spectral properties of the in-plane propagation
$S(r-r',t-t')$.
%
%averaged over ``the detuning fluctuations'' of the electromagnetic field. 
%In the adiabatic approximation, all fast processes and scattering are accounted for in the in-plane
%propagation probability $S(r-r',t-t')$, while the slow (low frequency)  and long-wavelength modes
%``detune'' the c-axis hopping without effecting the in-plane motion.
%The separation of the tunneling probability and the convolution of
%the in-plane Green function can be done on the basis of the adiabatic approximation
%(time scales separation). The long-wavelength ``detuning'' fluctuations ($\lambda > l_{mfp}$)
%can suppress the tunneling probability, but they should not change in-plane Green function.
%---------Analysis of P(r,t)
%The physical meaning of the function $P(r-r',t-t')$ is the probability to tunnel
%from one plane at the point $r$ and the time $t$ and tunnel back at the point $r'$ and the time $t'$.
The understanding of the properties of the tunneling probability $P(r-r',t-t')$,
describing  the effect of the ``detuning fluctuations'', is imperative for
any particular problem of the tunneling. 
The importance of this correlation function was first described in Ref.\cite{nazarov}.
The novel element here is the discussion of the spatial dependence of 
the tunneling probability function 
$ P(r-r',t-t')$. The spatial dependence appears to be very important for many questions
of the c-axis transport properties. As mentioned above, 
in previous studies of c-axis transport in cuprates,
 specific  properties of the tunneling
probability (e.g. the tunneling with or without the conservation 
of the in-plane momentum $k_\|$) were assumed. 
%we call the tunneling to be ``local''
%if the in-plane momentum is not conserved, or the probability of the tunneling is
%significant only if $k_F |r-r'| \sim< 1$. 
Here we analyze and calculate the tunneling probability from the fluctuation
spectrum of the electromagnetic field.
We can call the tunneling  ``diffusive''
if the in-plane momentum is not conserved
(if the momentum is conserved, it can be called specular).
In other words, the tunneling is diffusive, if the tunneling probability
$ P(r-r',t-t')$ is significant  
only if $  |r-r'|/l \stackrel{<}{\sim} 1 $
(where $l$ is a short length scale of order of a lattice constant). 
It should be noted that, 
generally speaking,
the question of the conservation of in-plane momentum during tunneling is another aspect
of tunneling not equivalent to the question of coherence or incoherence of tunneling
(that is the question of the dephasing of an electron).
Equation \ref{eq:current} can be rewritten in the following form\cite{Barone}:

\bleq
\beq
I(V)=
\frac{2e S t_\perp^2}{\hbar} \int dE dE' dk dk' A_1(k,E) A_2(k',E') ( f(E)(1-f(E'))  
P(E+eV-E',k-k')- && \n \\
 -f(E')(1-f(E))P(E'-eV-E,k-k') ) && \label{eq:big},
\eeq
\eleq\noindent
where $A_{1,2}(k,E)$ are the spectral functions, and  $f(E)$ is a Fermi function.

{\it The tunneling probability.}
We need to calculate the correlation function:
$\left< e^{i\hat{\varphi} (r,t)} e^{-i\hat{\varphi} (r',t')} \right>$. The averaging can be
done if we assume the field $ \varphi(r,\tau)$ is Gaussian correlated.
Using gauge invariance, the phase $ \varphi(r,\tau)$ can be rewritten as 
$ \varphi(r,\tau)=\int^\tau_{-\infty} \delta V(r,t) dt$, where $\delta V(r,t)$ is 
the local voltage difference between two planes.
This way we get an expression for the tunneling probability
(the calculation is a generalization of the exact calculation
from Ref.(\cite{Mahan}, p.273-277)):

\bleq
\beq
&& P(\delta r=r-r',\delta t=t-t')\equiv exp(- R(\delta r,\delta t)), \n \\ 
&& R(\delta r,\delta t) \equiv \int \frac{d\omega}{\omega^2}
 \int d^2q<\delta V_{q,\omega}^2>
 (1-cos(\omega \delta t + \vec{q} \vec{\delta r})) coth\frac{\omega}{2T}. \label{eq:tun-prob}
\eeq
\eleq\noindent
%where we have used the relation between the inter-plane phase  and  voltage difference:
% $\varphi_\ko=\delta V_\ko/i\omega$.

In this paper, we assume that the most effective ``detuning'' fluctuations are the voltage
fluctuations (or related charge fluctuations).
It is important to point out that the same method can be used to calculate the ``blocking''
of the tunneling due to any mechanism of  in-plane scattering. 
Since the nature of the ground state of cuprates and therefore the spectrum of 
the fluctuations
is not known, later we examine several general forms of the spectrum.
The problem of incoherent tunneling between a couple of two-dimensional planes is 
a natural generalization of the spin-boson model of quantum dissipation.

%The exchange part of the Coulomb interaction at long-wavelengths ($q \rightarrow 0$)
%is accounted for in the following below calculation. The role of the exchange interaction
%around $q\rightarrow (\pi,\pi)$ (spin-fluctuations theory) for ``dephasing'' 
%can be investigated using the phenomenological form of spin susceptibility
%(is it relevant ? outside the scope of the method?). 
%  Johnson-Nyquist noise. 
%Here for simplicity we assume that the fluctuations are uncorrelated 
%in two planes\cite{fluct-cor}.
% Most generally due to the fluctuation-dissipation theorem, 
% the spectrum of the voltage fluctuations 
% can be related with the imaginary part of the susceptibility. 

In the view of the importance of the spatial dependence of the tunneling probability,
we give {\it several different arguments} proving the diffusive nature of the tunneling
(if $\tau_{hop} \gg \tau_{ab}$) in the normal state.
First of all {\it a qualitative argument}: if the hopping time is much longer the in-plane
scattering time, an electron experiences many inelastic and elastic scattering processes
(both not conserving the direction of the in-plane momentum) 
before the hopping between planes.
Thus it is intuitively natural to think
that the momentum is not conserved after the hopping.
%(if the short transversal time is irrelevant for this consideration).
{\it A straightforward quantitative argument} is given by the analysis of the function 
$R(\delta r,\delta t)$
%(for simplicity we consider the $\delta t=0$ case, the analysis of divergences
% is also valid for $t \neq 0$)
 in the exponent of the
expression  for the tunneling probability (Eqn. \ref{eq:tun-prob}).
To separate the spatial and time dependence of the function $R(\delta r,\delta t)$,
we rewrite the multiplier in the integral as
$(1-cos(\omega \delta t + q_x \delta r))= (1-cos(\omega \delta t))+
cos(\omega \delta t)(1-cos(q_x \delta r)) -sin(q_x \delta r)sin(\omega \delta t)$.
The integral with the last term vanishes, because this term makes the integral expression
antisymmetric with respect to integration over $q_x$.
Thus, we can write 

\bleq
\beq
R(\delta r,\delta t)=R_0(\delta r,\delta t)+R_1(\delta r,\delta t), \n \\
R_0(\delta t)=
\int \frac{d\omega}{\omega^2}
\int dq_x q_y <\delta V_{q,\omega}^2> coth\frac{\omega}{2T}
(1-cos(\omega \delta t)), \n \\ 
 R_1(\delta r,\delta t)=
2\int \frac{d\omega}{\omega^2}
\int dq_x q_y <\delta V_{q,\omega}^2> coth\frac{\omega}{2T}
cos(\omega \delta t)sin^2((q_x \delta r)/2). \label{eq:r1}
\eeq
\eleq\noindent
The space-independent part $R_0(\delta t)$ is calculated later
in the paper (see Eqn. \ref{eq:ro}). Below we calculate the function
$R_1(\delta r,\delta t)$ which describes the spatial
dependence of the tunneling probability.
For this calculation we assume
that the fluctuations are uncorrelated in two planes (see below for 
a more general discussion),
and the interplane noise is just twice the in-plane Johnson-Nyquist (JN) noise
(voltage fluctuations).
The Johnson-Nyquist noise can be calculated from the spectral density of
Coulomb noise in the two-dimensional plane 
valid in the hydrodynamic approximation\cite{MT}:
\beq
<\delta V_{q,\omega}^2>\simeq 4\pi\sigma_Q \frac{\sigma_2 \omega}
{\omega^2+4\pi^2 \sigma_2^2q^2},
\label{eq:2d-Coulomb}
\eeq
where $\sigma_2$ is a two-dimensional conductance, and $\sigma_Q \equiv e^2/\hbar$.
We substitute Eqn.\ref{eq:2d-Coulomb} into the expression (\ref{eq:r1}) for
$R_1(\delta r,\delta t)$ and impose an upper cutoff $q_c$ on the $q$-integration
to take into account of the fact that the expression (\ref{eq:r1}) is strictly
valid only in the long-wavelength limit; thus we take $q \ll q_F$ (but still of
the general order of $q_F$). We also impose on the $\omega$-integration a lower
cutoff $\omega_l$, the choice of which will be discussed below. Then, taking into account
the fact that we are interested in values of $\delta t$ which are of the general order of
magnitude $\tau_{hop}$ and thus several orders of magnitude larger than $(\sigma_2 q_c)^{-1}$,
we see that to a good approximation $R_1(\delta r,\delta t)$ factorizes into a product
of a function of $\delta r$ and a function of $\delta t$:

\beq
R_1(\delta r,\delta t) \simeq F(\delta r) G(\delta t), \n \\
F(\delta r) \equiv \int_0^{2\pi}d\theta \int_0^{q_c \delta r}dx \frac{1-cos(xcos\theta)}{x}, \n \\
G(\delta t) \equiv \frac{2\sigma_Q}{\pi \sigma_2} \int_{\omega_l}^\infty
\frac{d \omega}{\omega} coth ( \frac{\omega}{2kT}) cos(\omega \delta t). \label{eq:gt}
\eeq
The expression $F(\delta r)$ is approximately $\frac{\pi}{8} (q_c\delta r)^2$ for
$q_c\delta r \ll 1$ and $2\pi ln(q_c\delta r)$ for $q_c\delta r \gg 1$. As for the
function $G(\delta t)$, we will see below that the ``interesting'' values of $\delta t$
(i.e. those for which the function $R_0(\delta t)$ does not suppress $P(\delta r, \delta t)$
too badly) are less or of order of $\hbar/2\pi\alpha kT$, where the dimensionless
quantity $\alpha$ is typically less or of order of $1$. Under these conditions,
provided $\hbar\omega_l \ll kT$ which will be satisfied by our choice of $\omega_l$ (cf. below),
the integral defining $G(\delta t)$ is dominated by its lower limit and approximately
given by the $\delta t$-independent expression

\beq
G(\delta t) \simeq \frac{2\sigma_Q}{\pi \sigma_2} \frac{kT}{\hbar \omega_l}.
\label{eq:gt-answer}
\eeq
We will make the choice $\omega_l \sim 1/\tau_{hop}$, on the grounds that
once we need to allow for appreciable interplane hopping, Eqn.\ref{eq:2d-Coulomb}
for the noise is no longer applicable (and we expect the expression for
$<\delta V^2_{q,\omega}>$ to decrease as a higher power of $\omega$ for $\omega \rightarrow 0$,
thereby effectively cutting off the integral (\ref{eq:gt})). We thus have
for $R_1(\delta r,\delta t)$  the approximate expression

\beq
R_1(\delta r,\delta t) \simeq (kT\tau_{hop}/\hbar)(\sigma_Q/\sigma_2)(2/\pi)F(q_c\delta r). \n
\eeq

The salient point, now, is that the quantity $kT\tau_{hop}/\hbar$, which is essentially
the ratio of the ab-plane and c-axis conductivities, is of order $10^2 -10^4$ for 
most of the cuprates, while the ratio $\sigma_2/\sigma_Q$ is never greater than about $10$.
Thus, the quantity $R_1$ has a value large compared to unity for values of $\delta r$
small compared to $1/q_c$, and we can approximate the expression $F(\delta r)$ by its
limiting form $\frac{\pi}{8} (q_c\delta r)^2$. Thus, the ``effective area'' 
$S_{eff}\equiv (\delta r_{eff})^2$ for which $R_1$ is appreciable is defined by

\beq
S_eff \sim \frac{4(\sigma_2/\sigma_Q)}{q_c^2 (kT\tau_{hop}/\hbar)} \n
\eeq
and by the above argument this is much smaller than $1/q_c^2$ and thus at most 
of the order of $1/q_F^2$. At distances of this order formulas such as (\ref{eq:2d-Coulomb})
should no longer be taken seriously, but the crucial upshot of the argument is that
{\it coherence between tunneling events separated in space by more than $\sim 1/q_F$ can be
simply neglected}. To put it differently, the tunneling is effectively local (diffusive);
the effective {\it rms} change in momentum in the course of a tunneling event is
of order of $q_F$ (cf. below). It is noteworthy that
this is so even if the momentum cutoff $q_c$ on the voltage fluctuations is only
a small fraction of $q_F$; consideration of shorter-wavelength fluctuations can
only strengthen this conclusion.

{\it Another argument estimates} directly the change of momentum in the process of tunneling.
The change of the momentum due to a fluctuation of the electromagnetic potential 
is $\delta p \sim (e\delta A)/c$, therefore
$ <\delta p^2>= (e/c)^2 <\delta A^2> $.
We can relate the correlation function of the vector potential with the correlation function
 of the scalar potential by gauge transformation (assuming only longitudinal fluctuations):
$ <A_{q,\omega}^2>=\frac{c^2q^2}{\omega^2}<\delta V_{q,\omega}^2>$.
% \[ <\varphi_{q,\omega}^2>=(\frac{2\pi}{q})^2 \frac{Dq^2}{\omega^2+ (Dq^2(2\pi/q))^2} \]
Using Eqn. \ref{eq:2d-Coulomb} at low frequencies $\omega <  \sigma_2 q$, 
the correlation function can be approximated as
$ <A_{q,\omega}^2>= \frac{4\pi\sigma_2 c^2}{\omega^2} \frac{q^2}
                    {\omega^2+\sigma_2^2 q^2} \simeq \frac{4\pi c^2}{\omega^2 \sigma_2} $. 
Thus the variance of momentum $<\delta p^2>$ is
$
<\delta p^2>= \frac{e^2}{\hbar c^2} \int dq~ q 
   \int d\omega \frac{c^2}{\omega^2}
   \frac{\omega coth\frac{\omega}{2T}}{\sigma_2}.
$ 
We see that the integral over frequency is diverging for $\omega < T$ as
$\int \frac{d\omega}{\omega^2}$. This integral can be cut off again 
on $1/\tau_{hop}$, therefore
$<\delta p^2> \sim q_c^2 \frac{\sigma_Q}{\sigma_2} (\tau_{hop} kT)$ 
%\sim k_F^2 \frac{\sigma_Q}{\sigma_2}\frac{\tau_{hop}}{\tau_{ab}}$ 
(at any finite temperature).
 This estimate gives a result equivalent
to the earlier calculation.
% At $T=0K$, $<\delta p^2> \sim k_F^2 \frac{\sigma_Q}{\sigma_2}$ with logarithmic accuracy. 
 This indicates again that the in-plane momentum is
completely randomized after the process of tunneling. 
All these arguments validate theories of c-axis transport in the normal state
assuming non-conservation of in-plane momentum during tunneling.
%The process of the tunneling
%can be viewed as completely inelastic with respect to the initial momentum and energy 
%(if $\alpha \ll 1$ the energy is quasi-conserved ?).
The fact of the non-conservation of momentum $k_\|$  
in the normal state (if $\tau_{hop} \gg \tau_{ab}$)
is quite general, a sufficient
condition as can be seen from the above discussion is the ohmic density of the noise
$<\delta V_r(\omega)^2> \sim \omega$ for $\omega \rightarrow 0$.
It is important to stress that the ``diffusivity'' of the tunneling is 
due to specific form of the spectrum
of the voltage fluctuations (and not due to short links or impurities!),
for instance, it may not be true in 
the superconducting state.
It is important to point out that the question of the
in-plane momentum conservation for c-axis tunneling
can be examined experimentally\cite{hussey} supporting or disproving the above arguments.

% the discussion of \alpha and local tunneling probability

When the tunneling is diffusive, the ``detuning fluctuations'' are simply
the local voltage fluctuations 

\beq
 <\delta V_{\omega}^2>= \int d^2 q <\delta V_{q,\omega}^2> \equiv \alpha
             \omega , \n
\eeq
where $\alpha$ is the microscopic parameter describing the ohmic density of the noise.
As can be seen below this microscopic parameter $\alpha$ 
is sufficient to describe all {\it dc} and {\it ac}
dependencies of the c-axis conductivity. 
When the two planes are widely separated and isolated 
(the situation possibly realized in $Bi-2201$),
the noise spectra in each plane are uncorrelated, so that the interplane noise spectrum
is just the sum of the noise spectra in each plane. In such a case, 
the coefficient $\alpha$ should be
determined only by the properties of the copper-oxygen plane. If the planes are 
moved closer together,
so that the inter-plane Coulomb interaction become relevant,
the intensity of the interplane noise (assuming no interplane hopping) increases due to
the presence of the acoustic (out-of-phase) plasmon in this bi-layer structure.   
Unfortunately, realistically the noise between planes can be suppressed  and 
correlated at low frequencies
because of interplane hopping
and become dependent on the interlayer structure thus implying different values of $\alpha$
for different cuprate materials. It is known experimentally that the detailed temperature
dependence of the c-axis resistivity is very sensitive to several factors (sample preparation,
interlayer structure and doping). The role of the interlayer structure 
(different intercalating atoms,
chains and additional layers present in some compounds) and the structure of the tunneling
matrix element $t_\perp (k_x,k_y)$ (e.g. in-plane anisotropy) is not clear, it makes
the question of the interlayer noise and the c-axis transport properties even more complex.

% additional paragraph with discussion of
% S_{eff} and
% the full spectrum of noise
% discussion of \alpha

We have examined several models describing the voltage noise to estimate 
the microscopic parameter
$\alpha$ (see the appendix for the discussion of this question, also \cite{shieh}). 
The goal of such exercise  is to verify a crude consistency 
between the estimate of
$\alpha$ from microscopic noise and the parameter $\alpha$ required 
to describe the c-axis transport dependencies.
The difficulty
(not surprising since these calculations assume Fermi-liquid or 
diffusive spectra of density fluctuations)
 common to all of the calculations (see the appendix) is that 
the ``microscopic'' value of $\alpha$
is significantly smaller than the value of order of $1$ necessary to 
explain the c-axis transport properties.
For our approach to be valid a large ($\alpha \sim 1$) density of ``detuning fluctuations''
(not present in RPA or Fermi liquid pictures) is vital. 
{\it An alternative approach} is to extract the charge fluctuation noise directly from the optical 
reflectivity measurements. The most dramatic
difference between good metals and cuprates seen in Raman and optical measurements 
of the in-plane dielectric constant $\epsilon_{ab}(q,\omega)$ is 
that the low-frequency noise 
% (temperature independent in the normal state)
 for cuprates is ohmic (linear) even for $q\rightarrow 0$ .
If we write $Im(-1/\epsilon_{ab}(q,\omega))=\gamma \omega$ for $q \simeq 0$, then 
from the experimental data\cite{timusk}
$\gamma \simeq 0.2 eV^{-1}$. Since 
$<\delta V_{\omega}^2>=\int \frac{dq q}{(2\pi)^2} V_2(q) Im(-1/\ep) 
\simeq \frac{e^2 q_{c}}{2\pi} \gamma \omega$, 
we get
$\alpha \simeq \frac{e^2 q_{c}}{2\pi} \gamma$. 
If we take the upper cut-off wave vector $q_c \sim \frac{2\pi}{a}$
($a$ is a in-plane lattice constant), we get $\alpha \simeq \frac{e^2}{a} \gamma \simeq 1.2$.
It shows that $\alpha$ may be of order of $1$, 
exactly what is required to explain the c-axis transport properties. 
We now use the ``local'' approximation justified above to calculate the
{\it dc} and {\it ac} c-axis conductivity.

%The main point here is to show that $\alpha$ of order of 1.  
% discussion of P(E)
% Below we summarize the results of detailed calculations based
% on Eqns. \ref{eq:current} and \ref{eq:tun-prob}.

First of all, we  calculate the local tunneling probability $P(r=0,t)$ 
(or rather its Fourier transform $P(\epsilon, k-k')$).
% and use it later for calculations of the c-axis conductivity. 
The quantity $P(\epsilon, k-k')$ can be interpreted as a probability
to exchange an energy $\epsilon$ and momentum $(k-k')$ with fluctuating
fields.
Since, as argued above, the tunneling probability is strongly peaked 
at $r=0$ (if $\tau_{hop} \gg \tau_{ab}$),
 the Fourier transform in momentum space
$P(\epsilon,k-k')$ is  essentially independent of $(k-k')$, therefore we omit 
the index $(k-k')$ below.
For the small values of the coefficient $\alpha$ ($\alpha \ll 1$) describing
the spectral density of the local voltage noise 
 $<\delta V_{\omega}^2>= \alpha \omega $,
the tunneling probability can be calculated analytically.
The function $R_0(\delta t)$ in the exponent of the $P(\delta r=0,\delta t) $
is 

\beq
&& R_0(\delta t)=\int\frac{d\omega}{\omega^2}<\delta V_{\omega}^2>
(1-cos(\omega \delta t)) coth(\frac{\omega}{2kT})= \n \\
&&=\int_0^\infty \frac{d\omega}{\omega^2} \alpha \omega
(1-cos(\omega \delta t)) coth(\frac{\omega}{2kT}) \simeq \n \\
&& \simeq \frac{\delta t}{\tau_\phi}+2\alpha ln(\omega_c/kT), \label{eq:ro}
\eeq
where $\tau_\phi=1/2\pi\alpha kT$ and $\omega_c$ is a high-frequency cutoff
which we associate with the
inverse of the transversal time $\hbar/\tau_{tr} \sim 1eV$.
After a Fourier transform, it gives
the probability of tunneling 
%in the process
%creating an excitation of  energy $\epsilon$ is
$P(\epsilon)= S_{eff}(\frac{kT}{\omega_c})^{2\alpha}
\frac{2\pi\alpha kT}{\epsilon^2+(2\pi\alpha kT)^2}$,
where $S_{eff}$ is the effective area of tunneling discussed above.
% (which is of order $1/k_F^2$). 1/l^2
For $\alpha$ of order of 1, the tunneling probability is strongly suppressed
and weakly depends on $\epsilon$ (for $\epsilon \leq kT$).
 It cannot be calculated explicitly analytically,
but at small $\epsilon$ can be approximated as a function independent of $\epsilon$:  
$P(\epsilon) \simeq S_{eff}(\frac{kT}{\omega_c})^{2\alpha}\frac{1}{2\pi\alpha kT}$.

%The function $I(r=0,t)$ is given by
%$ I(r=0,t)= \int \frac{d\omega}{\omega^2} \alpha\omega coth\frac{\omega}{2T}
%(1-cos\omega t) \sim const +2(\alpha T)t+\alpha ln(\omega_c/2T)$.
%Therefore, the tunneling probability is
%$P(E=0)= \int_0^\infty d\tau T^{2 \alpha} e^{-2 \pi \alpha k T \tau}= 
%T^{2 \alpha}/2 \pi \alpha k T $.
% \cdot \tau^{out}_{\phi}$
%, where $ \tau^{out}_{\phi}=1/2 \pi \alpha k T $.
 
% conductivity

{\it The optical conductivity $\sigma_c(\omega, kT)$. The tunneling conductance 
$\sigma_c(V)$.}
Equation \ref{eq:big} can be further transformed assuming 
the diffusive tunneling probability.
In this case, we can integrate over the in-plane momenta ($k, k'$) separately.
The next transformation is due to the detailed balance condition
(see  Ref.\cite{nazarov}),
eventually we can write
the expression for the {\it dc} conductivity $\sigma_c(V)$
as a function of the applied voltage $V$:  
%After integrating over in-plane momenta 
%\beq
%&& I(V)=
%2e S t_\perp^2 \int dE dE' \nu_L(E) \nu_R(E') P(E-eV-E')   \n \\
%&&f(E)(1-f(E'))(1-e^{\beta V}) 
%\eeq
%Using the detailed balance relation for $P(E)$, the final expression

\beq
\sigma_c(V)=\frac{e t_\perp^2 d \nu_{2D}^2}{\hbar }\frac{1-e^{-\beta eV}}{V}
%\frac{I(V) d}{V S} \sim 
\int_{-\infty}^{+\infty} d\epsilon \frac{\epsilon P(eV-\epsilon)}
{1-e^{-\beta \epsilon}}. \label{eq:last}
\eeq
From Eqn. \ref{eq:last} in the limit of small voltage ($V \ll kT$)
(for all $\alpha$ as long as the tunneling is diffusive)
 appropriate
for the {\it dc}  measurements, we calculate the temperature dependence of 
the c-axis conductivity

\beq
\sigma_c(T)=\frac{e^2}{\hbar} t_\perp^2 d \nu_{2D}^2 S_{eff}
\left(\frac{kT}{\omega_c}\right)^{2\alpha}\frac{1}{2\pi\alpha}.
\label{eq:result}
\eeq 
This result describes the c-axis resistivity $\rho_c(T)$ either constant or diverging
at low temperatures found experimentally in several compounds 
($Bi$-family, $Hg-1201$, $Tl-2212$,$Tl-1212$ and slightly underdoped 
$La-214$). 

%K+11: more details

In order to calculate the optical conductivity $\sigma_c(\omega)$, we make 
the following observation. If we apply the external {\it dc} voltage, the tunneling
probability acquires an additional phase factor
$P(t-t') \sim exp(\frac{i}{\hbar} \int_{t'}^t V d\tau)$ (in Eqn.\ref{eq:big}, it
gives a shift in the energy difference (after a Fourier transform) $P(E+eV-E')$).
Thus schematically, the conductance is
$\sigma_c(V) \sim \frac{1}{V}\int dt' e^{\frac{i}{\hbar}V(t-t')} ( ... )$.
For the {\it ac} voltage,

\beq 
P(t-t') \sim exp(\frac{i}{\hbar} \int_{t'}^t V e^{i\omega \tau} d\tau)= \n \\
=exp(\frac{i}{\hbar}V \frac{e^{i\omega t}-e^{i\omega t'}}{i\omega}) \approx \n \\
\approx 1+\frac{V}{\hbar}\frac{e^{i\omega t}}{\omega}(1-e^{i\omega (t-t')}).
\eeq
In the linear response (the optical conductivity is a linear response) and
separating a corresponding harmonic of the current proportional to $e^{i\omega t}$,
we conclude that
the dependence of the conductivity $\sigma_c(\omega)$
 on the frequency $\omega$
is equivalent to the dependence on the voltage $V$, 
such that $\sigma_c(\omega)=\sigma_c(V \rightarrow \omega)$
(if the tunneling is incoherent and diffusive).
Namely,

\beq
\sigma_c(\omega)=\frac{e t_\perp^2 d \nu_{2D}^2}{\hbar }\frac{1-e^{-\beta \omega}}{\omega}
%\frac{I(V) d}{V S} \sim 
\int_{-\infty}^{+\infty} d\epsilon \frac{\epsilon P(\omega-\epsilon)}
{1-e^{-\beta \epsilon}}. \label{eq:freq-sigma}
\eeq

A ubiquitous nearly flat optical response for $\sigma_c (\omega)$ is
observed experimentally (if measurements exist) in these compounds.
Indeed, a qualitative and numerical analysis of the  frequency dependence 
$\sigma_c(\omega)$ of Eqn.\ref{eq:freq-sigma} (or equivalently dependence 
on the voltage) indicates a very weak dependence 
% (or even the absence of dependence) 
on frequency.

It is interesting that under the conditions of incoherent tunneling  ohmic
$I-V$ curves (constant conductance $\sigma_c(V)$ as a function of voltage) correspond
to a flat optical conductivity $\sigma_c(\omega)$.
The correspondence $\sigma_c(\omega)=\sigma_c(V \rightarrow \omega)$  can be directly checked 
experimentally by comparing the tunneling conductance $\sigma_c(V)=I(V)/V$  and the frequency 
dependence of $\sigma_c(\omega)$
from the optical reflectivity measurements. 
The graphs of  $\sigma_c(V)$ and $\sigma_c(\omega)$ can be taken from experimental papers
\cite{Suzuki}.
It appears to be roughly true 
for optimally doped and underdoped compounds
in the normal state.

% Two channels picture

{\it Discussion of experiments.}
The above results are insufficient to describe all experimental data for different dopings.
In optimally doped $La-214$ and $YBaCuO$ and other overdoped cuprates the c-axis
resistivity has a linear temperature dependence with a large ``residual'' value
(intercept at $T=0K$) \cite{thallium}.
In these compounds the hopping time becomes comparable to the in-plane scattering time.
Thus we can assume that specular tunneling  becomes possible.
In this crossover situation between two limiting pictures of the tunneling between
planes and the anisotropic band along c-axis, we can think about two channels
of conduction.
One channel is diffusive, while another one is specular.
%We may think about two scenarios how the second channel can appear.
%First, if the properties around the in-plane fermi surface are isotropic,
The tunneling probability is the sum of probabilities to tunnel
without and with the conservation of the in-plane momentum. In this case,
the total c-axis conductivity is the sum of conductivities in each channel.
If the fermi surface has very anisotropic properties, electrons from
one part of the fermi surface can tunnel specularly, while electrons from other parts
tunnel diffusively.
In this scenario, the second channel of conduction 
(conserving $k_\|$) can be due to the diagonal parts
of the fermi-surface in the normal state. It appears from photoemission experiments that 
the quasiparticles along diagonals of the Brillouin zone ($k_x=\pm k_y$) have 
longer life-times $\tau_{ab,diag}$ (which should be compared with the hopping time).
These quasiparticles can tunnel then with conserved momentum. 
It is natural
to suggest that for overdoped cuprates the c-axis transport is dominated by
incoherent, but specular channel, while for underdoped cuprates
the diffusive channel is only present.

%Since
%these channels dissipate in parallel, we should add resistivities 
%associated with each channel.
We can calculate the conductivity of a specular channel, if
we assume that the time dependence of the specular tunneling probability
is $P(\delta t) \simeq exp(-\frac{\delta t}{\tau_\phi})$ with $\tau_\phi$
calculated for a weak detuning, that is $\tau_\phi=1/2\pi\alpha kT$.
Thus we substitute the tunneling probability of the form
$ P_2(\epsilon-\epsilon',k-k')= \delta (k-k') 
\frac{2\pi\alpha kT}{(\epsilon-\epsilon')^2+(2\pi\alpha kT)^2}$
(or any form 
$ P_2(\epsilon-\epsilon',k-k') \simeq \delta (k-k') 
\frac{1}{kT} f(\frac{\epsilon-\epsilon'}{kT})$) 
to the Eqn. \ref{eq:big}, we get the contribution to the c-axis conductivity

\beq
\sigma_{2,c}(T)=\frac{e^2}{\hbar} t_\perp^2 d \nu_{2D}
\frac{A}{k T},
\eeq
where $A$ is a numerical coefficient.
This result is a well-known result for incoherent tunneling with conservation
of the in-plane momentum \cite{kumar}. It is suggested to explain the linear
temperature dependence of c-axis resistivity observed in 
some compounds \cite{thermal}. Another important consideration is
that the band calculations predict a significant angular dependence
of $t_\perp$ in some families of cuprates\cite{Xiang}. Due to this reason,
 the contribution
from specular channel (from diagonal parts of the fermi-surface) can be reduced.

%In some situations, the large residual resistivity (due to a diffusive channel)
% and the linear temperature dependence (due to a specular channel)
% are observed together implying the presence of both conducting channels simultaneously.
%The picture of two channel conduction can explain all variety
%of different c-axis temperature dependencies observed in cuprates. 

Yet another complexity of cuprates with multiple planes per unit cell is
the question of intra-unit cell and inter-cell conduction.
In this case, the resistance associated with hopping between planes of
the unit cell (intra-cell resistance) and the resistance associated with hopping
between different unit cells (inter-cell resistance) should be discussed.
The total resistance is certainly a sum of intra- and inter-unit cell resistances.
It may not be correct to assume (as is frequently done in the literature)
that the intra-cell resistance is negligible; a systematic experimental
investigation of this question is necessary. We hope to discuss
this question elsewhere.

In conclusion, a picture of c-axis interlayer (and inter-cell) tunneling
strongly suppressed by voltage fluctuations is proposed.
This approach can provide a consistent understanding of
observed temperature and frequency dependencies of c-axis conductivity
in the normal state.

%We leave other cases ($Tl-2201$, overdoped $Y Ba_2 Cu_3 O_7$ and $La_{1.7}Sr_{0.3}CuO_4$) 

%$\sigma_c(T) \sim T^{2\alpha-1}/\alpha$. For $\alpha \ll 1$ the temperature dependence of
%the {\it dc} resistivity is almost linear which is the case for optimally doped or overdoped
%$Y Ba_2 Cu_3 O_7$ and $La_{2-x}Sr_xCuO_4$ compounds. For $\alpha \sim 1$ the temperature
%dependence of resistivity is ``semiconducting-like''.
%At certain crossover temperature $T^*$ in these materials (underdoped)
% the resistivity always becomes linear (while diverging below  $T^*$). 

%\footnote{ check if detailed symmetry can be used here too,
%as well as explicitly detailed symmetry in expression for $P(E)$}:

%For the case of $\alpha \ll 1$ , we get Drude-like response.
%Again, it is in good correspondence with optimally doped $Y Ba_2 Cu_3 O_7$. 
% It would be interesting to verify this result for Tl(2212) ( to my knowledge, there are no yet
% measurements with subtracted phonon background of $\sigma_c (\omega)$ for this material).
% the $\rho_c$ is $T$, but flat $\sigma_c(\omega)$.

%It is important to stress that the same model
%gives Drude-like and flat optical conductivity as a function of a single parameter $\alpha$.
%While in the literature, it is believed to be a real crossover from one regime to another.

% Conclusion of fluctuation induced local tunneling
% compare with  local tunneling in disorder systems (the argument ??)
% Discussion, Comparison of theory and experiment
% more detailed discussion of experiments

% Conclusion

This work was supported by the National Science Foundation through
the Science and Technology Center for Superconductivity (grant no. DMR-91-20000)
and through grant no. DMR-99-86199.
We thank A. Shnirman and L.-Y. Shieh for useful discussions.

{\it Appendix.}

In this appendix, we demonstrate an example of a calculation of the parameter
$\alpha$ describing the low frequency voltage fluctuations
based on a particular microscopic model of the density fluctuation spectrum.
At the end of the appendix, we list several results calculated for various other
microscopic models.

If the two planes are widely separated, then the voltage fluctuations can be 
treated independently in each plane. In this case, the interplane noise is just twice 
the in-plane local Johnson-Nyquist noise in  each plane.
For a high-density electron gas in a hydrodynamic approximation
($\omega \tau \ll 1$ and $ql \ll 1$), the charge density-density susceptibility
of the two-dimensional electron gas
can be written as\cite{Fetter}

\beq
\chi(k,\omega)=\frac{s^2 k_{TF} k^2}{2\pi e^2}
\frac{1}{\omega(\omega+i/\tau)-s^2k^2-s^2 k_{TF} k}, \n
\eeq
where $s^2=v_F^2/2$, $k_{TF}=\frac{2\pi n_0 e^2}{ms^2}$ is the Thomas-Fermi wave number,
and $\tau$ is a phenomenological relaxation time.
The spectral density of the voltage fluctuations is

\beq
<V_{k,\omega}^2>=V_k^2 Im \chi(k,\omega), \n 
\eeq
where $V_k=\frac{2\pi e^2}{k}$ is the two-dimensional Coulomb interaction.
Eventually, we need to calculate the partial frequency-dependent spectral density
of the voltage noise $<V_\omega^2>=\int \frac{dk k}{2\pi}<V_{k,\omega}^2>$.
The calculation gives

\beq
<V_\omega^2>=\int \frac{dk k}{2\pi} \frac{(2\pi e^2)^2}{k^2}\frac{s^2k_{TF}k^2}{2\pi e^2} &&\n \\
\frac{\omega (1/\tau)}{(\omega/\tau)^2 +(s^2k^2+s^2k_{TF}k)^2}= &&\n \\
\simeq \frac{e^2}{s^2k_{TF}\tau} \omega ln(\frac{s^2k_{TF}^2 \tau}{\omega_c}), &&\n
\eeq
where $\omega_c$ is an infra-red cut-off frequency.
It implies to the accuracy of the value of the logarithm that the parameter $\alpha$ is

\beq
\alpha \equiv \frac{e^2}{s^2k_{TF}\tau} \frac{1}{\hbar}=\frac{\hbar}{2\epsilon_F \tau}. \n
\eeq   

We can rewrite  this expression in the following form:

\beq
\alpha=\frac{1}{\pi} \frac{\sigma_Q}{\sigma_{2D}}, \n 
\eeq
where $\sigma_Q=e^2/\hbar$, 
$\sigma_{2D}=e^2 \nu_{2D} D=\frac{2}{\pi}\frac{e^2 (\epsilon_F \tau)}{\hbar^2}$ is 
the two-dimensional conductivity (it can be shown to have
a Drude frequency dependence). 
The above results may be used to estimate the value of the parameter
$\alpha$ in the cuprates. Two difficulties can be seen from such literal
application of the above model. First of all, the value of $\alpha$ is significantly
smaller than one. Second, and more importantly, the parameter $\alpha$ appears
to be temperature dependent. It should be realized that the spectrum
of charge fluctuations in the cuprates is much more complex and not represented correctly
by the above simple model. We investigated several other simple microscopic models
in order to get further insight into this question.
At this moment, it seems more reasonable
to extract the charge fluctuation spectrum directly from experiment as shown in the text
of this paper.

%The parameter $\alpha$ calculated from the JN noise (Eqn.\ref{eq:2d-Coulomb}) is equal to
%(assuming $2\pi\sigma_2/L_{charact} > \omega$)
%$\alpha=\frac{\sigma_Q}{2\pi^3 \sigma_2} ln\frac{2\pi\sigma_2}{\omega L}$. 
% If we assume $\sigma_2$ as the temperature dependent conductivity, we get inconsistency.

%We  should think about the conductance $\sigma_2$ as the quantity describing  the {\it local} 
%JN noise (not the temperature dependent in-plane conductance). The temperature
%independent spectrum of the noise in the normal state is seen in experiments 
%(see the discussion in the text).
The results for several other microscopic models are summarized below.
If the spectrum of charge fluctuations is given by the weakly damped
acoustic two-dimensional plasmon, then the parameter $\alpha$ is
$\alpha \sim \frac{\sigma_Q^2}{s^2(1+k_{TF}d)}$, where $d$ is the inter-plane distance.
%(in-phase and out-of-phase) or by the diffusive modes,
%we get $\alpha=\frac{\hbar}{2 \epsilon_F \tau}$ (where $\tau$ is a lifetime of plasmon
%or the elastic scattering time).
Another calculation taking the voltage noise due to electron-hole pairs
of the two-dimensional Fermi liquids
% calculated in the RPA approximation 
gives $\alpha \simeq 1/4\pi$.
% (remarkably universal answer). 
If the interplane Coulomb interaction is taken
into account in the RPA approximation for the same calculation (el-hole pairs), 
$\alpha=1/(2\pi k_F d)$.
%The weakly damped two-dimensional plasmons have a superohmic density of states, and thus do not
%make the tunneling local
%(the probability of tunneling is a power-law of the distance between points). 
We hope to present the details of calculations and expanded arguments elsewhere.

\end{multicols}

%A possible explanation for a linear {\it correction} (except $Tl-2201$) to the resistivity:
%if for some reasons  there is a probability to tunnel
%with conserved momentum $k_\|$, then the correction $\delta \sigma_c (T) \sim \nu_{2D}/kT$,
%$\delta \rho_c (T) \sim kT$ (due to $1/\tau_{ab} \sim kT$).
% The value of $\rho_c (T=0)$ is not zero (not due to impurities) in all cases. 
%It could be thought as a temperature
%dependent $S_{eff}$. Another possible reason (especially at high temperatures ($La-214$))
%is the thermal expansion  of c-axis interplane distance ( meaning the reduction of $t_\perp$).


\begin{references}
\bibitem{hop}
The estimates of the hopping time $\tau_{hop} $ for different materials as well as
the inverse of the hopping time $\hbar/\tau_{hop}$ which is a measure of kinetic energy
associated with the hopping:
$\tau_{hop} \simeq  10^{-11} s, \hbar/\tau_{hop} \simeq 1K$
                           (optimally doped $Hg-1201$ and $Tl-2201$),
$\tau_{hop} \simeq 2* 10^{-12} s, \hbar/\tau_{hop} \simeq 5K$  ($Tl-1212$),
$\tau_{hop} \simeq 0.5* 10^{-10} s, \hbar/\tau_{hop} \simeq 0.2K$  ($Tl-2212$),
$\tau_{hop} \simeq 4* 10^{-10} s,\hbar/\tau_{hop} \simeq 0.025K $  ($Bi-2212$ and $Bi-2201$),
$\tau_{hop} \simeq 1.5*10^{-12} s, \hbar/\tau_{hop} \simeq 6.7K, $ 
(optimally doped $LaSrCuO$),
$\tau_{hop} \simeq 4*10^{-14} s, \hbar/\tau_{hop} \simeq 250K$   (overdoped $LaSrCuO$),
$\tau_{hop} \simeq 6*10^{-14} s, \hbar/\tau_{hop} \simeq 167K $  (optimally doped $YBaCuO$). 
For comparison, at
$T=100 K$ the in-plane scattering time is $\tau_{ab} = \hbar /kT \simeq 10^{-13} s$.
The traversal time under the barrier is very short of order of
$10^{-15}s$ and should be distinguished from the hopping time.
\bibitem{thallium}
The case of $Tl-2201$ appears to be very special, because the c-axis resistivity
has a strong linear temperature dependence with a small ``residual'' value. 
It can be contrasted with the c-axis resistivity of a very similar single-layer
compound $Hg-1201$ (identical to $Tl-2201$ in many respects). While the absolute
values of resistivity at $T_c$ are the same in both materials, the c-axis resistivity
of $Hg-1201$ has a diverging upturn in its temperature dependence.
\end{references}
\end{document}